\documentclass[aps,prd,superscriptaddress,twocolumn,secnumarabic]{revtex4-1}
\usepackage{amsmath}

\pdfoutput=1

\usepackage{bm}
\usepackage{times}
\usepackage{braket}
\usepackage{color,graphicx}
\usepackage[utf8]{inputenc}
\usepackage[T1]{fontenc}
\newcommand{\cb}{{\cal{B}}}
\newcommand{\bea}{\begin{eqnarray}}
\newcommand{\eea}{\end{eqnarray}}
\newcommand{\e}[1]{\,\mathrm{#1}}
\newcommand{\E}[1]{\times 10^{#1}}

\usepackage[normalem]{ulem}

\definecolor{nicered}{rgb}{0.7,0.1,0.1}
\definecolor{nicegreen}{rgb}{0.1,0.5,0.1}

\usepackage{hyperref}
\hypersetup{colorlinks,citecolor=nicegreen,linkcolor=nicered}
\hypersetup{colorlinks=true}

\begin{document}
\title{Scalar leptoquarks from GUT to accommodate the
$B$-physics anomalies}

\author{Damir Be\v cirevi\' c} \email[Electronic address:]{damir.becirevic@th.u-psud.fr}
\affiliation{Laboratoire de Physique Th\' eorique (B\^at.\ 210), Universit\' e Paris-Sud and CNRS, F-91405 Orsay-Cedex, France}

\author{Ilja Dor\v sner} \email[Electronic address:]{dorsner@fesb.hr}
\affiliation{University of Split, Faculty of Electrical Engineering, Mechanical Engineering and Naval Architecture in Split (FESB), Ru\dj era Bo\v skovi\' ca 32, 21 000 Split, Croatia}

\author{Svjetlana Fajfer} \email[Electronic address:]{svjetlana.fajfer@ijs.si}
\affiliation{Department of Physics, University of Ljubljana, Jadranska 19, 1000 Ljubljana, Slovenia}
\affiliation{Jo\v zef Stefan Institute, Jamova 39, P.\ O.\ Box 3000, 1001
  Ljubljana, Slovenia}

\author{Darius A.\ Faroughy} \email[Electronic address:]{darius.faroughy@ijs.si}
\affiliation{Jo\v zef Stefan Institute, Jamova 39, P.\ O.\ Box 3000, 1001
  Ljubljana, Slovenia}

\author{Nejc Ko\v snik} \email[Electronic address:]{nejc.kosnik@ijs.si}
\affiliation{Department of Physics, University of Ljubljana, Jadranska 19, 1000 Ljubljana, Slovenia}
\affiliation{Jo\v zef Stefan Institute, Jamova 39, P.\ O.\ Box 3000, 1001
  Ljubljana, Slovenia}
  
\author{Olcyr Sumensari} \email[Electronic address:]{olcyr.sumensari@pd.infn.it}
\affiliation{Dipartimento di Fisica e Astronomia ``G.\ Galilei'', Universit\` a di Padova, Italy}
\affiliation{Istituto Nazionale Fisica Nucleare, Sezione di Padova, I-35131 Padova, Italy}  
  
\begin{abstract}
We address the $B$-physics anomalies within a two scalar leptoquark model.
The low-energy flavor structure of our set-up originates from two $SU(5)$ operators that relate Yukawa couplings of the two leptoquarks. The proposed scenario has a UV completion, can accommodate all measured lepton flavor universality ratios in $B$-meson decays, is consistent with related flavor observables, and is compatible with direct searches at the LHC. We provide prospects for future discoveries of the two light leptoquarks at the LHC and predict several yet-to-be-measured flavor observables.
\end{abstract}
\maketitle

\section{Introduction}
\label{sec:intro}
Lepton flavor universality~(LFU) ratios appear to be very interesting observables to test the validity of the Standard Model~(SM).
Several experiments found that LFU ratios
$R_{D^{(*)}} = { \cb( B \to D^{(*)} \tau \bar\nu_\tau)}/{\cb( B \to
  D^{(*)}  l \bar\nu_l)}$, $l= e,\mu$, are larger than
$R_{D^{(*)}}^\mathrm{SM}$. The measurements of
$R_{D}$~\cite{Lees:2013uzd,Hirose:2016wfn,Aaij:2015yra} differ
by $\sim 2\,\sigma$ with respect to the SM prediction~\cite{Lattice:2015rga} and by $\sim 3\,\sigma$ in the case of $R_{D^{\ast}}$~\cite{Fajfer:2012vx,Bernlochner:2017jka,Bigi:2017njr}. Further
hints of LFU violation in transitions $b \to c \ell \nu$ ($\ell = \mu, \tau$) were observed in
$R_{J/\psi}$ ratio between $B_c \to J/\psi \ell \nu$ decay widths~\cite{Aaij:2017tyk}. On the other hand, the LHCb
experiment has measured LFU ratios
$R_{K^{(\ast)}} = {\cb (B\to K^{(\ast)} \mu\mu)}/{\cb ( B\to
  K^{(\ast)} e e)}$ related to the neutral-current process
$b\to s l l$ and found them to be systematically lower than expected
in the SM.  While $R_K$ was measured in a single
kinematical region, $q^2 \in [1.1,6]\e{GeV}^2$~\cite{Aaij:2014ora},
$R_{K^*}$ was measured also in the region
$q^2\in [0.045,1.1]\e{GeV}^2$~\cite{Aaij:2017vbb}. The three measured
$R_{K^{(*)}}$ deviate from the SM predictions at
$\sim 2.5\,\sigma$ level~\cite{Hiller:2003js,Bordone:2016gaq}.

New Physics~(NP) explanations of the $B$-physics anomalies suggest a presence
of one or more TeV scale mediators which couple to left-handed currents with
predominantly third generation
fermions~\cite{Bordone:2016gaq,Buttazzo:2017ixm,Bhattacharya:2014wla,Feruglio:2016gvd,Hiller:2014yaa}. Among
the most prominent NP candidates are leptoquarks (LQs). 
It turns out that a single scalar LQ cannot provide solution to the both $B$-physics anomalies. The scalar LQ that can explain $R_{K^\ast}$ is not suitable for accommodating  $R_{D^\ast}$ and vice versa. 
By using an 
effective theory approach, it was shown in
Ref.~\cite{Buttazzo:2017ixm} that of all possible single mediators
only one particular vector LQ can generate suitable $V\!-\!A$
operators for the anomalies and satisfy both low-energy and high-$p_T$
constraints. The construction of ultra-violet (UV) complete models for
these scenarios became a challenge that was addressed in
Refs.~\cite{Assad:2017iib,Bordone:2018nbg,Bordone:2017bld,Greljo:2018tuh,DiLuzio:2017vat,Blanke:2018sro,Calibbi:2017qbu}. Another
possible approach is to consider models with several mediators. The low
energy $V\!-\!A$ structure can be generated by integrating out
two scalar
LQs~\cite{Buttazzo:2017ixm,Crivellin:2017zlb,Marzocca:2018wcf}. Incidentally, one can
explore two scalar LQs even when they are known to generate operators
with Lorentz structures other than $V\!-\!A$ for
$R_{D^{(*)}}$.

In this Letter we propose a UV complete model based on $SU(5)$ Grand Unified Theory~(GUT) with two light scalar LQs that can address both anomalies. The LQs in question are $R_2(\mathbf{3},\mathbf{2},7/6)$ and $S_3(\overline{\mathbf{3}},\mathbf{3},1/3)$, where we specify their representation under the SM gauge group $SU(3)_c \times SU(2)_L \times U(1)_Y$. At low energies, $R_2$ generates a combination of scalar and tensor effective operators that accommodate $R_{D^{(*)}}$, while $S_3$ generates a $V\!-\!A$ operator which accommodates $R_{K^{(*)}}$. In our set-up, since the Yukawa interactions have a common $SU(5)$ origin, both LQs share one Yukawa matrix. If we take into account all relevant flavor constraints we find that the preferred region in the parameter space is compatible with direct searches at the LHC. Furthermore, if we demand perturbativity of all the couplings to the GUT scale, we find that the mass of $R_2$ needs to be around $1\e{TeV}$. In the following we present our set-up, outline the UV completion, discuss the low-energy phenomenology and LHC signatures.

\section{Set-up}

\label{sec:setup}
The interactions of $R_2$ and $S_3$ with the SM fermions are
\begin{equation}
\label{eq:one}
\mathcal{L}  \supset Y_{R}^{ij} \bar{Q}^\prime_i \ell^\prime_{Rj} R_2+ Y_L^{ij} \bar{u}^\prime_{Ri} \widetilde{R}_2^\dagger L^\prime_j +Y^{ij} \bar{Q}^{\prime C}_{i} i \tau_2 ( \tau_k S^k_3) L^\prime_{j},
\end{equation}
where $Y_L$, $Y_R$, and $Y$ are Yukawa matrices, $\tau_k$ denote the Pauli matrices, $S_3^k$ are the $SU(2)_L$ triplet components, $\widetilde{R_2} \equiv i \tau_2 R_2^\ast$, and $i,j,k=1,2,3$. We omit hermitian conjugate parts throughout the Letter. This part of the lagrangian, in the mass eigenstate basis, reads
\begin{align}
\label{eq:two}
\begin{split}
\mathcal{L} \supset 
&+(V Y_R E_R^\dagger)^{ij} \bar{u}_{Li}\ell_{Rj}R_2^{\frac{5}{3}} + (Y_R E_R^\dagger)^{ij} \bar{d}_{Li}\ell_{Rj} R_2^{\frac{2}{3}}\\
&+(U_R Y_L U)^{ij} \bar{u}_{Ri} \nu_{Lj} R_2^{\frac{2}{3}}- (U_R Y_L)^{ij} \bar{u}_{Ri}\ell_{Lj} R_2^{\frac{5}{3}}\\
&-(Y U)^{ij} \bar{d}^C_{Li} \nu_{Lj} S_3^{\frac{1}{3}} +2^\frac{1}{2}(V^* Y U)^{ij} \bar{u}^C_{Li} \nu_{Lj} S_3^{-\frac{2}{3}}\\
&- 2^\frac{1}{2} Y^{ij} \bar{d}^C_{Li} \ell_{Lj} S_3^{\frac{4}{3}} -(V^* Y)^{ij} \bar{u}^C_{Li} \ell_{Lj} S_3^{\frac{1}{3}},
\end{split} 
\end{align}
where $R_2^{(Q)}$ and $S_3^{(Q)}$ are the charge (and mass)
eigenstates with charge $Q$. We define the mass eigenstates
$u_{L,R}=U_{L,R} u^\prime_{L,R}$, $d_{L,R}=D_{L,R} d^\prime_{L,R}$,
$\ell_{L,R}=E_{L,R} \ell^\prime_{L,R}$, and
$\nu_{L}=N_L \nu^\prime_{L}$, where
$U_{L,R}$,
$D_{L,R}$, $E_{L,R}$, and $N_L$ are unitary matrices.
$V= U_L D_L^\dagger \equiv U_L$ and
$U\equiv E_L N_L^\dagger \equiv N_L^\dagger$ are the CKM and PMNS
matrices, respectively.

We adopt the following features for the Yukawa matrices
\begin{equation}
\label{eq:yL-yR-y}
Y_R E_R^\dagger=(Y_R E_R^\dagger)^T\,, \qquad Y=-Y_L\,,
\end{equation}
and assume 
\begin{equation}
\label{eq:yL-yR}
Y_R E_R^\dagger = \begin{pmatrix}
0 & 0 & 0\\ 
0 & 0 & 0\\ 
0 & 0 & y_R^{b\tau}
\end{pmatrix},~ 
U_R Y_L = \begin{pmatrix}
0 & 0 & 0\\ 
0 & y_L^{c\mu} & y_L^{c\tau}\\ 
0 & 0 & 0
\end{pmatrix}\,,
\end{equation}
where $U_R^{22}=\cos \theta \equiv c_\theta$, $U_R^{23}=-\sin \theta \equiv -s_\theta$, and $|U_R^{11}|=1$. Relevant NP parameters are $m_{R_2}$, $m_{S_3}$, $y_R^{b\tau}$, $y_L^{c\mu}$, $y_L^{c\tau}$, and $\theta$.

\paragraph*{$SU(5)$ embedding}
In the simplest $SU(5)$ scenario that can accommodate light $R_2$ and $S_3$ and (re)produce the associated flavor structure of Eqs.~\eqref{eq:yL-yR-y} and~\eqref{eq:yL-yR}, the scalar sector needs to contain $\underline{\bm{45}}$ and $ \underline{\bm{50}}$ whereas the SM fermions comprise $\overline{\bm{5}}_{i}$ and $\bm{10}_i$, where $i(=1,2,3)$ is a generation index. We omit the $SU(5)$ indices and underline scalar representations throughout this section.

To generate all three operators of Eq.~\eqref{eq:one} it is sufficient to introduce $a^{ij} \bm{10}_i \overline{\bm{5}}_j \overline{\underline{\bm{45}}}$, and $b^{ij} \bm{10}_i \bm{10}_j  \underline{\bm{50}}$, where $a$ and $b(=b^T)$ are $3 \times 3$ matrices in generation space. The former contraction couples $R_2 \in \underline{\bm{45}}$ ($S_3 \in \underline{\bm{45}}$) with the right-handed up-type quarks (quark doublets) and leptonic doublets, while the latter generates couplings of $R_2 \in \underline{\bm{50}}$ with the quark doublets and right-handed charged leptons. To break $SU(5)$ down to the SM gauge group we can use $\underline{\bm{24}}$~\cite{Georgi:1974sy} or $\underline{\bm{75}}$~\cite{Hubsch:1984pg,Hubsch:1984qi}. This allows us to write either $m\, \underline{\bm{45}} \, \overline{\underline{\bm{50}}} \, \underline{\bm{24}}$ or $m\, \underline{\bm{45}} \, \overline{\underline{\bm{50}}} \, \underline{\bm{75}}$, where $m$ is a dimensionful parameter. The two $R_2$'s that reside in $\underline{\bm{45}}$ and $\underline{\bm{50}}$ mix through either of these two contractions allowing us to end up with two light scalars, i.e., $R_2$ and $S_3$, and one heavy $R_2$ state that completely decouples from the low-energy spectrum for large values of $m$. The relevant lagrangian after the $SU(5)$ breaking, in the mass eigenstate basis of the two light LQs, is 
\begin{align}
\label{eq:three}
\begin{split}
\mathcal{L} \supset &+s_\phi (V^\prime b E_R^{\prime \dagger})^{ij} \bar{u}_{Li}\ell_{Rj}R_2^{\frac{5}{3}} + s_\phi (b E_R^{\prime \dagger})^{ij} \bar{d}_{Li}\ell_{Rj} R_2^{\frac{2}{3}}\\
& +c_\phi (U^\prime_R a U^\prime)^{ij} \bar{u}_{Ri} \nu_{Lj} R_2^{\frac{2}{3}}-c_\phi (U^\prime_R a)^{ij} \bar{u}_{Ri}\ell_{Lj} R_2^{\frac{5}{3}}\\
&+2^{-\frac{1}{2}} (a U^\prime)^{ij} \bar{d}^C_{Li} \nu_{Lj} S_3^{\frac{1}{3}}-(V^{\prime *} a U^\prime)^{ij} \bar{u}_{Li}^C \nu_{Lj} S_3^{-\frac{2}{3}} \\
&+ a^{ij} \bar{d}_{Li}^C \ell_{Lj} S_3^{\frac{4}{3}}+2^{-\frac{1}{2}} (V^{\prime *}  a)^{ij} \bar{u}^C_{Li} \ell_{Lj} S_3^{\frac{1}{3}},
\end{split} 
\end{align}
where we define the mixing angle between the two $R_2$'s to be $\phi$. The primed unitary transformations in Eq.~\eqref{eq:three}, i.e., $V^\prime$, $E_R^\prime$, $U^\prime_R$, $U^\prime$, as well as Yukawa matrices $a$ and $b$ are defined at the GUT scale. It is now trivial to compare Eq.~\eqref{eq:two} with Eq.~\eqref{eq:three} to obtain the following identification after renormalization group running from the GUT scale down to the electroweak scale: $a \rightarrow - \sqrt{2} Y$, $c_\phi U^\prime_R a \rightarrow U_R Y_{L}$, $ s_\phi b E_R^{\prime \dagger} \rightarrow Y_R E_R^\dagger$, $V^\prime \rightarrow V$, and $U^\prime \rightarrow U$. 
Our particular ansatz given in Eqs.~\eqref{eq:yL-yR-y} and~\eqref{eq:yL-yR} is consistent with this identification if we take both $R_2$ states to mix maximally, i.e., $s_\phi=c_\phi=1/\sqrt{2}$. Clearly, the two $SU(5)$ operators proportional to $a$ and $b$ suffice to generate the three operators of Eq.~\eqref{eq:one} associated with the Yukawa matrices $Y$, $Y_L$, and $Y_R$.

\paragraph*{Perturbativity} We advocate the case that the low-energy Yukawa couplings have an $SU(5)$ origin. We implement the low-energy lagrangian of Eq.~\eqref{eq:one} in SARAH-4.12.3~\cite{Staub:2013tta} and obtain one- and two-loop beta function coefficients to accomplish the renormalization group running from the electroweak to the GUT scale which we set at $5 \times 10^{15}$\,GeV. The low-energy Yukawas that we use as input are the ones presented in Eq.~\eqref{eq:yL-yR} and we scan over $y^{b\tau}_R$, $y^{c\mu}_L$, and $y^{c\tau}_L$ to identify the region of parameter space for which all Yukawa couplings remain below $\sqrt{4\pi}$ up to the GUT scale. We find, for example, that the most relevant Yukawa coupling contributions for the running of $y^{b\tau}_R$ are
\begin{equation*}
  16 \pi^2 \frac{d \ln y^{b\tau}_R}{d \ln \mu}=|y^{c\mu}_L|^2+|y^{c\tau}_L|^2+\frac{9}{2}|y^{b\tau}_R|^2+\frac{1}{2}y^{2}_t +\dots,
\end{equation*}
where $y_t$ is the top Yukawa coupling. 

\paragraph*{Proton decay} LQs are commonly associated with proton decay. It is thus important to address the issue of matter stability. $R_2$ cannot mediate proton decay at tree-level either directly or through mixing via one or two Higgs scalars in our set-up. It is an innocuous field with regard to the issue of matter stability. It can also be arranged that $S_3$ does not contribute towards proton decay. One prerequisite for this to happen is the absence (or suppression) of the contraction $c^{ij} \bm{10}_i \bm{10}_j \underline{\bm{45}}$ that couples $S_3$ to two quark doublets~\cite{Dorsner:2012nq}. The other prerequisite is that $S_3$ does not mix with any other LQ with diquark couplings. Both prerequisites can be simultaneously satisfied in a generic $SU(5)$ framework~\cite{Dorsner:2017wwn}. It is thus possible to have light $R_2$ and $S_3$ without any conflict with the stringent experimental limits on matter stability.

\section{Low-energy phenomenology}

\paragraph*{Charged-current decays}
The relevant effective lagrangian for (semi-)leptonic decays is
\begin{equation}
\begin{split}
{\cal L}^{d \to u \ell \bar \nu}_{\mathrm{eff}} = -\frac{4 \, G_F}{\sqrt{2}} &V_{ud}\big[  (1+g_{V_L}) 
(\bar{u}_L \gamma_\mu d_L)(\bar{\ell}_L \gamma^\mu \nu_{L})\\ 
&+ g_{S_L}(\mu)\, (\bar{u}_R  d_L)(\bar{\ell}_R \nu_{L}) \\
&+ g_T(\mu)\, (\bar{u}_R  \sigma_{\mu \nu} d_L) (\bar{\ell}_{R} \sigma^{\mu \nu}\nu_L)
\big]\,,
\label{eq:hamiltonian-semilep}
\end{split}
\end{equation}
where neutrinos are in the flavor basis. The effective Wilson coefficients of $d\to u \ell \bar{\nu}_{\ell^\prime}$ are related to the LQ couplings at the matching scale $\Lambda (\approx 1$\,TeV) via the expressions
\begin{equation}
\label{eq:semilep-WC}
\begin{split}
g_{S_L}(\Lambda) = 4 \, g_T(\Lambda) &= \frac{y_L^{u\ell^\prime}\, {y_R^{d\ell}}^{\ast}}{4 \sqrt{2} \, m_{R_2}^2 \, G_F V_{ud}}\,,\\
g_{V_L}  &= -\dfrac{y^{d\ell^\prime} \left(V y^\ast\right)^{u\ell}}{4\sqrt{2} \, m_{S_3}^2 G_F V_{ud}}\,.
\end{split}  
\end{equation}
From the above equations we learn that the only transitions affected by $R_2$ in our scenario are $b\to c\tau \bar{\nu}_\ell$. $S_3$, on the other hand, contributes to processes involving $u$, $c$, $s$, $b$, and $\ell,\ell^\prime=\mu,\tau$, but gives a negligibly small contribution to $R_{D^{(\ast)}}$. 

To compute $R_D$ we employ the $B\to D$ form factors calculated using the lattice QCD~\cite{Na:2015kha,Lattice:2015rga}, resulting in prediction $R_D^{\mathrm{SM}}=0.29(1)$ which is $\approx 2\,\sigma$ below the experimental average $R_D^{\mathrm{exp}}=0.41(5)$~\cite{Lees:2012xj,Huschle:2015rga,Aaij:2017deq}. On the other hand, the $B\to D^\ast$ form factors have never been computed on the lattice at nonzero recoil. Thus, for $R_{D^\ast}$ we consider the leading form factors extracted from the $B\to D^\ast l \bar{\nu}$ ($l=e,\mu$) spectra~\cite{Amhis:2016xyh}, which are combined with the ratios $A_0(q^2)/A_1(q^2)$ and $T_{1-3}(q^2)/A_1(q^2)$ computed in Ref.~\cite{Bernlochner:2017jka}. We obtain the value $R_{D^\ast}^{\mathrm{SM}}=0.257(3)$ which is $\approx 3\,\sigma$ below the experimental average $R_{D^\ast}^{\mathrm{exp}}=0.30(2)$~\cite{Amhis:2016xyh}. Moreover, to confront the scalar (tensor) effective coefficients in Eq.~\eqref{eq:semilep-WC} with low-energy data, we account for the SM running from the matching scale $\mu = \Lambda$ down to $\mu=m_b$, while the vector coefficient is not renormalized by QCD~\cite{Gonzalez-Alonso:2017iyc}.

We include in the fit several (semi-)leptonic decays which are sensitive to the $S_3$ couplings~\cite{Dorsner:2017ufx}. Particularly, the LFU ratios $R_{D^{(\ast)}}^{\mu/e} = \mathcal{B}(B\to D^{(\ast)}\mu \bar{\nu})/\mathcal{B}(B\to D^{(\ast)} e \bar{\nu})$~\cite{Glattauer:2015teq,Abdesselam:2017kjf} impose severe constraints to simultaneous explanations of the $b\to c$ and $b\to s$ anomalies~\cite{Becirevic:2016oho}. Furthermore, we consider $\mathcal{B}(B\to \tau \bar{\nu})$ and the kaon LFU ratio $R^K_{e/\mu}= \Gamma(K^-\to e^- \bar{\nu})/\Gamma(K^-\to \mu^- \bar{\nu})$~\cite{Cirigliano:2007xi}, both in agreement with their SM predictions. See, for example, Ref.~\cite{Dorsner:2017ufx} for further discussion.

\paragraph*{Neutral-current decays} The standard left-handed effective Hamiltonian for the $b\to s$ (semi-)leptonic transition can be written as
\begin{equation}
\mathcal{H}^{b\to s l l}_{\mathrm{eff}} = -\dfrac{4 G_F \lambda_t }{\sqrt{2}}  \sum_{i=7,9,10} C_i(\mu)\mathcal{O}_i(\mu)\,,
\end{equation}
where $\lambda_t = V_{tb}V_{ts}^\ast$. The relevant operators for our discussion are
\begin{equation}
  \mathcal{O}_{9(10)} = \dfrac{e^2}{(4\pi)^2} \,\big{(}\bar{s}_L\gamma_\mu  b_L\big{)} \big{(}\bar{l}\gamma^\mu (\gamma^5)l \big{)}\,.
\end{equation}
In our set-up, only $S_3$ contributes at tree-level via~\cite{Dorsner:2017ufx}
\begin{equation}
\begin{split}
  \delta C_9^{\mu\mu} = - \delta C_{10}^{\mu\mu} &= \dfrac{\pi v^2}{ \lambda_t \alpha_{\mathrm{em}}} \dfrac{y^{b\mu} \big{(}y^{s\mu}\big{)}^\ast}{m_{S_3}^2}\\
  &=\dfrac{\pi v^2}{ \lambda_t \alpha_{\mathrm{em}}} \frac{\sin 2\theta\, (y_L^{c\mu})^2}{2 m_{S_3}^2}\,.
\end{split}
\end{equation}
In the second line we explicitly write the dependence of $\delta C_9^{\mu\mu}$ on $\sin 2\theta$. This angle allows one to vary $R_{K^{(*)}}$, independently of $R_{D^{(*)}}$. Since we consider a scenario with relatively small Yukawa couplings, it is a very good approximation to neglect loop-induced contributions of $R_2$ (and $S_3$) to this transition. For a different set-up, see Ref.~\cite{Becirevic:2017jtw}. The $1\,\sigma$ interval  $C_9^{\mu\mu}=-C_{10}^{\mu\mu} \in (-0.85,-0.50)$ is obtained by performing a fit to the clean $b\to s l l$ observables, namely, $R_K$, $R_{K^{\ast}}$, and $\mathcal{B}(B_s\to \mu\mu)$~\cite{DAmico:2017mtc,Capdevila:2017bsm}.

Contributions to the left-handed current operators in $b\to s l l$ transition unavoidably imply contributions to $B\to K^{(\ast)}\nu\nu$ decays which are well constrained by experiments. These decays are governed by
\begin{equation}
\mathcal{L}_{\mathrm{eff}}^{b\to s\nu\nu}= \dfrac{\sqrt{2}G_F \alpha_{\mathrm{em}} \lambda_t}{\pi} ~C_L^{ij}~\big{(}\bar{s}_L\gamma_\mu b_L\big{)}\big{(}\bar{\nu}_{L\,i} \gamma^\mu \nu_{L\,j} \big{)}\,,
\end{equation}
where $C_L^{ij}=\delta_{ij}C_{L}^{\mathrm{SM}}+\delta C_L^{ij}$ is the Wilson coefficient which includes the SM contribution $C_{L}^{\mathrm{SM}}=-6.38(6)$~\cite{Altmannshofer:2009ma} and the contribution $\delta C_L^{ij}$ from NP. Similarly as in the $b\to s l l$ transition, the only tree-level contribution to $b\to s \nu\bar{\nu}$ comes from the $S_3$ state and reads~\cite{Dorsner:2017ufx}
\begin{equation}
\delta C_L^{ij} = \dfrac{\pi v^2}{2\alpha_{\mathrm{em}} \lambda_t} \dfrac{y^{bj} \big{(}y^{si}\big{)}^\ast}{m_{S_3}^2}\,,\quad i,j=\mu,\tau\,.
\end{equation}
These effective coefficients modify the ratios $R_{\nu\nu}^{(\ast)}=\mathcal{B}(B\to K^{(\ast)}\nu\nu)/\mathcal{B}(B\to K^{(\ast)}\nu\nu)^{\mathrm{SM}}$ in the following way
\begin{equation}
\label{eq:Rnunu}
R_{\nu\nu}^{(\ast)} = \dfrac{\sum_{ij}|\delta_{ij}C_L^{SM}+\delta C_{L}^{ij}|^2}{3|C_L^{\mathrm{SM}}|^2}.
\end{equation}
In Sec.~\ref{sec:num} we confront the predictions of $R_{\nu\nu}^{(\ast)}$ with experimental bounds $R_{\nu\nu}<3.9$ and $R_{\nu\nu}^{\ast}<2.7$~\cite{Grygier:2017tzo}.

\paragraph*{Further flavor constraints}
Our low-energy fit also includes constraints which will be more
extensively discussed in a future publication. These are (i) the
$B_s-\bar{B}_s$ mixing amplitude, which is shifted by the $S_3$
box-diagram, proportional to 
$\sin^2 2\theta \left[(y_L^{c\mu})^2 + (y_L^{c\tau})^2\right]^2/m_{S_3}^2$, (ii)
the experimental limits 
$\mathcal{B}(\tau\to\mu\phi) \sim \cos^4 \theta (y_L^{c\mu} y_L^{c\tau})^2/m_{S_3}^4$, which is bounded to remain below  $8.4\times 10^{-8}$, and
$\mathcal{B}(\tau\to\mu\gamma)^{\mathrm{exp}}<4.4\times
10^{-8}$~\cite{Patrignani:2016xqp}, (iii) the muon $g-2$, which shows
$\approx 3.6\,\sigma$ discrepancy with respect to the
SM~\cite{Beringer:1900zz} but receives only small contribution in our
set-up, and (iv) the $Z$-boson couplings to leptons measured at
LEP~\cite{ALEPH:2005ab}, which are modified at loop level by both
$R_2$ and $S_3$.  Finally, we have also checked that our model is
compatible with measured $D-\bar D$ mixing parameters.

\section{Numerical results}
\label{sec:num}

\begin{figure}[!b]
  \centering
  \includegraphics[scale=0.4]{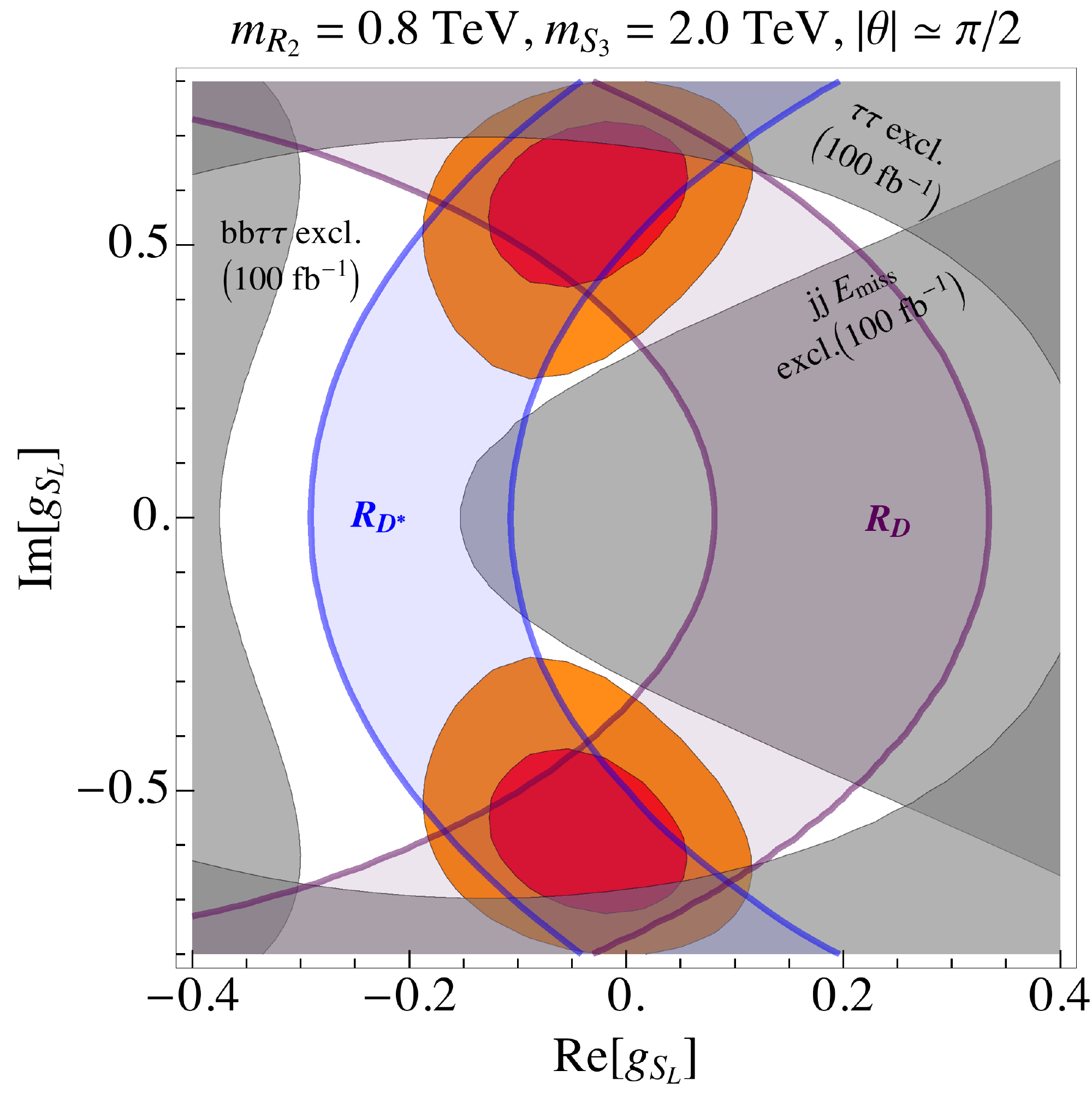}
  \caption{Results of the flavor fit in the $g_{S_L}$ plane, as defined in Eq.~\ref{eq:hamiltonian-semilep} for the transition $b\to c\tau \bar{\nu}_\tau$. The allowed $1\,\sigma (2\,\sigma)$ regions are rendered in red (orange). Separate constraints from $R_D$ and $R_{D^\ast}$ to $2\,\sigma$ accuracy are shown by the blue and purple regions, respectively. The LHC exclusions, as discussed in Sec.~\ref{sec:LHC}, are depicted by the gray regions.
  \label{fig:gSplotRDst}}
\end{figure}

\begin{figure}[!htbp]
  \includegraphics[scale=.62]{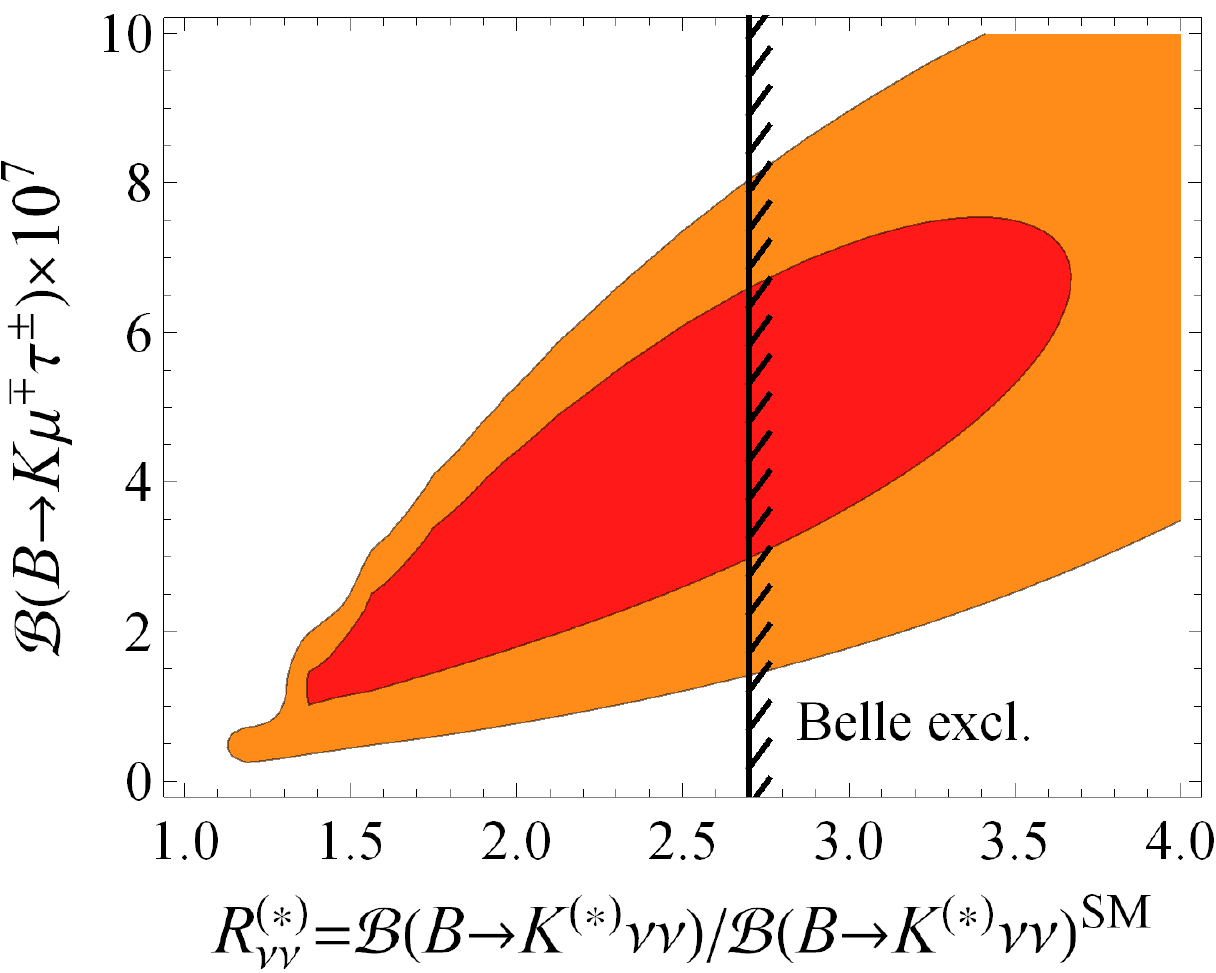}
  \caption{$\mathcal{B}(B\to K \mu \tau)$ is plotted against $R_{\nu\nu}=\mathcal{B}(B\to K^{(\ast)} \nu \bar{\nu})/\mathcal{B}(B\to K^{(\ast)} \nu \bar{\nu})^{\mathrm{SM}}$ for the $1\,\sigma$ (red) and $2\,\sigma$ (orange) regions of Fig.~\ref{fig:gSplotRDst}. The black line denotes the current experimental limit, $R_{\nu\nu}^{\ast}<2.7$~\cite{Grygier:2017tzo}.}
  \label{fig:prediction}
\end{figure}

We perform a fit to the observables listed above by varying the parameters $y_R^{b\tau}$, $y_L^{c\mu}$, $y_L^{c\tau}$ and $\theta$, which were introduced in Sec.~\ref{sec:setup}. The fit requires $S_3$ to be more massive than $R_2$. The masses $m_{R_2}$ and $m_{S_3}$ are set to the lowest values allowed by projected LHC constraints, namely, $m_{R_2} = 800$\,GeV and $m_{S_3} = 2$\,TeV, as we discuss later on. Note that in our flavor fit we obtain two solutions corresponding to small ($\theta \sim 0$) and large ($|\theta| \sim \pi/2$) mixing angles. These two solutions successfully suppress the key constraints, such as $R_{K^{(*)}}$ and $\Delta m_s$ since they are proportional to $\sin 2\theta$.  Further inclusion of $\mathcal{B}(\tau\to\mu\phi)\propto \cos^4\theta$ in the fit selects the solution with $|\theta| \approx \pi/2$ as the only viable one. The results of our fit in the $g_{S_L}$ complex plane are shown in Fig.~\ref{fig:gSplotRDst} to $1\,\sigma$ and $2\,\sigma$ accuracies. The SM point is excluded with $3.8\,\sigma$ significance, while the best fit point provides a perfect agreement with $R_{D^{(\ast)}}$ and $R_{K^{(\ast)}}$. Interestingly, a simultaneous explanation of $R_D$ and $R_{D^\ast}$ requires complex $g_{S_L}$, which is why we consider complex $y_R^{b\tau}$~\cite{Sakaki:2014sea,Becirevic:2018uab}. Note that the phase in $y_R^{b\tau}$ causes no observable CP violating effects. The best fit point is consistent with the LHC constraints superimposed on the same plot. A purely imaginary solution is:
\begin{equation}
\label{eq:Nejc}
  \mathrm{Re} [g_{S_L}] = 0,~
  |\mathrm{Im} [g_{S_L}]| = 0.59
     \left(^{+0.13}_{-0.14}\right)_{1\,\sigma} \left(^{+0.20}_{-0.29}\right)_{2\,\sigma}\,.
\end{equation}
\noindent An important prediction of our scenario is that $\mathcal{B}(B\to K\mu\tau)$ is bounded from above and below, as illustrated in Fig.~\ref{fig:prediction}. At $1\,\sigma$ we obtain
\begin{equation}
1.1 \times 10^{-7}\lesssim  \mathcal{B}( B\to K \mu^\pm \tau^\mp) \lesssim 6.5\times 10^{-7} \,.
\end{equation}
This value is smaller than the current $\mathcal{B}(B\to K \mu \tau)^{\mathrm{exp}}<4.8\times 10^{-5}$~\cite{Lees:2012zz}, which can certainly be improved by LHCb and Belle-II. Note that our prediction can easily be translated into similar modes via relations $\mathcal{B}(B\to K^\ast \mu\tau)\approx 1.9\times \mathcal{B}(B\to K \mu \tau)$ and $\mathcal{B}(B_s\to \mu\tau)\approx  0.9 \times \mathcal{B}(B\to K \mu \tau)$~\cite{Becirevic:2016zri,Glashow:2014iga,Guadagnoli:2015nra}. Another important prediction of our set-up is a $\gtrsim 50\%$ enhancement of $\mathcal{B}(B\to K ^{(\ast)} \nu \nu)$, which can be tested in the near future at Belle-II. Remarkably, these two observables are highly correlated as depicted in Fig.~\ref{fig:prediction}. Furthermore, we predict a lower bound on $\mathcal{B}(\tau \to \mu \gamma)$, which lies just below the current experimental limit,
\begin{equation}
 \mathcal{B}(\tau \to \mu \gamma) \gtrsim 1.5\E{-8}\,.
\end{equation}

Finally, our description of the $B$-physics anomalies, and most particularly $R_{D^{(\ast)}}$, strongly depends  on the assumption that the LQ states are not too far from the TeV scale. Thus, these particles are necessarily accessible at the LHC, yielding also predictions for the direct searches which we discuss next.

\section{LHC PHENOMENOLOGY}
\label{sec:LHC}

\begin{figure}[t]
\begin{center}
 \includegraphics[width=0.9\hsize]{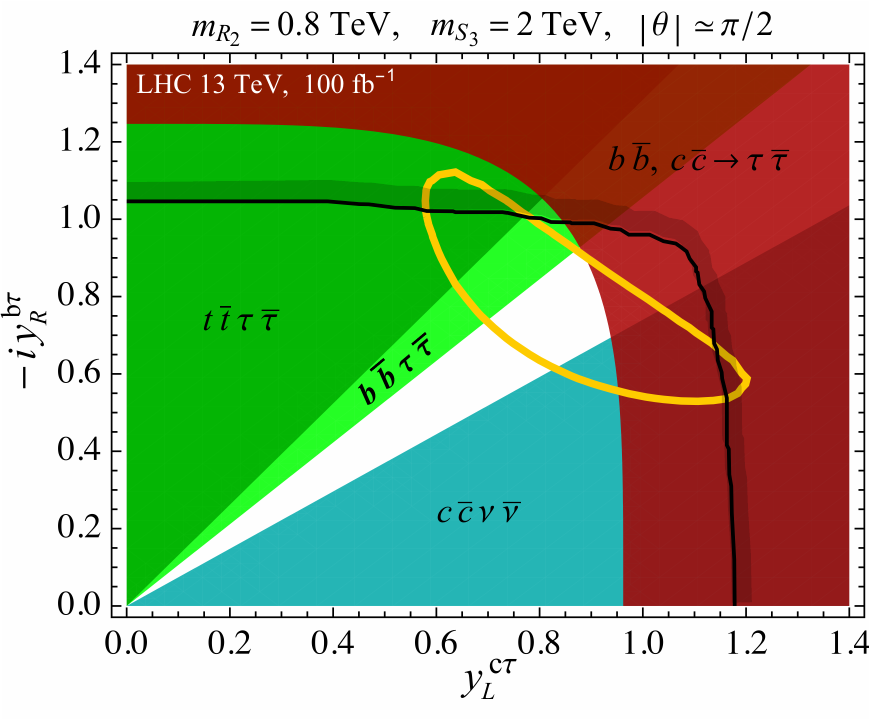}
 \vspace{-0.5cm}
\end{center}
\caption{Summary of the LHC limits for each LQ process at a projected luminosity of 100\,fb$^{-1}$ for $m_{R_2}=800$\,GeV, $m_{S_3}=2$\,TeV, and $|\theta| \approx\pi/2$. The red region corresponds to the exclusion limit from the high-$p_T$ di-tau search by ATLAS~\cite{Aaboud:2017sjh}, while the green and turquoise exclusion regions come from LQ pair production searches by CMS~\cite{Sirunyan:2017yrk,CMS:2017kmd,Sirunyan:2018nkj}. The region above the solid black contour represents values of the couplings that become non-perturbative at the GUT scale. The region inside the yellow contour corresponds to the $1\,\sigma$ fit to the low-energy observables. }
\label{fig:LHCbound}
\end{figure}

Direct searches at the LHC can play an important role in constraining LQ model(s) aiming to explain the $R_{D^{(*)}}$ and $R_{K^{(*)}}$ anomalies. In the following we show that the benchmark masses $m_{R_2}=800$\,GeV and $m_{S_3}=2$\,TeV are currently allowed by the high-$p_T$ and direct search experiments at the LHC and present exclusion limits for a projected LHC luminosity of 100\,fb$^{-1}$ of data. 

\paragraph*{High-$p_T$ di-tau tails} The dominant NP contributions to $q\bar q\!\to\!\tau\tau$ production, in view of the flavor structure of Eq.~\eqref{eq:yL-yR}, come from the $t$-channel exchange of $R_2^{\frac{5}{3}}$ and $R_2^{\frac{2}{3}}$ states in charm and bottom annihilation, respectively. Similar contributions from $S_3$ depend on the value of the mixing angle $\theta$. As discussed in Sec.~\ref{sec:num}, the low-energy fit prefers $|\theta| \approx \pi/2$. In this case an almost exact flavor alignment takes place between $\tau$ and the third quark generation, meaning that only the exchange of $S_3^{-\frac{1}{3}}$ from initial $b\bar b$ collisions contributes to $\tau\tau$ production. Following Ref.~\cite{Faroughy:2016osc}, we confront this scenario with data by recasting the most recent search by ATLAS~\cite{Aaboud:2017sjh} at 13\,TeV and 36.1\,fb$^{-1}$ for a $Z^\prime\!\to\!\tau_{\text{had}}\bar\tau_{\text{had}}$ heavy resonance in the high-mass tails. Our results for the 95\% C.L.\ limits in the $y_L^{c\tau}$--$(y_R^{b\tau}/i)$ plane are given by the red exclusion region in Fig.~\ref{fig:LHCbound} for the two benchmark masses, $|\theta| \approx\pi/2$, and the LHC luminosity of 100\,fb$^{-1}$. 

\paragraph*{Leptoquark pair production} For the benchmark masses, bounds from pair-produced LQs can only be derived for $R_2$. The dominant decay channels for each charged eigenstate are $R_2^{\frac{2}{3}} \to \tau b,\,\nu c$ and $R_2^{\frac{5}{3}} \to \tau t,\, \tau c$ with the corresponding branching fractions fixed by the squared ratio of Yukawa couplings $y_L^{c\tau}/y_R^{b\tau}$. To set limits on $gg \to (R_2^{\frac{2}{3}})^* R_2^{\frac{2}{3}}$, we use CMS results from the search~\cite{Sirunyan:2017yrk} for $b\bar b \tau\bar\tau$ final states and the multi-jet plus missing energy search~\cite{CMS:2017kmd} for $c\bar c \nu\bar\nu$ final states, i.e.,~$jj$ plus missing energy signature. The 95\% C.L.\ exclusion limits are given by the light green and turquoise regions in Fig.~\ref{fig:LHCbound} for a luminosity of 100\,fb$^{-1}$. As for the pair production of $R_2^{\frac{5}{3}}$ states, we employ the search by CMS~\cite{Sirunyan:2018nkj} targeting $t\bar t \tau\bar\tau$ final states. Results from this search are given by the dark green exclusion region in Fig.~\ref{fig:LHCbound}.

\section{Conclusions}
We propose a two scalar LQ extension of the SM that can accommodate all measured LFU ratios in $B$-meson decays and related flavor observables, while being compatible with direct search constraints at the LHC. The extension has an $SU(5)$ origin that relates Yukawa couplings of the two LQs through a mixing angle and all Yukawas remain perturbative up to the unification scale. We provide prospects for future discoveries of the two light LQs at the LHC and spell out predictions for several yet-to-be-measured flavor observables. In particular, we predict and correlate $\mathcal{B}(B\to K\mu\tau)$ with $\mathcal{B}(B\to K ^{(\ast)} \nu \nu)$. We also predict a lower bound for $\mathcal{B}(\tau\to\mu\gamma)$ which is just below the current experimental limit.

\begin{acknowledgments}
S.F., D.A.F., and N.K.\ acknowledge support of the Slovenian Research Agency under the core funding grant P1-0035. This work has been supported in part by Croatian Science Foundation under the project 7118 and the European Union's Horizon 2020 research and innovation programme under the Marie Sklodowska-Curie grant agreements N$^\circ$~674896 and 690575.
\end{acknowledgments}


\begin{thebibliography}{99}
\bibitem{Lees:2013uzd} 
  J.~P.~Lees {\it et al.} [BaBar Collaboration],
  Phys.\ Rev.\ D {\bf 88}, no. 7, 072012 (2013)
  [arXiv:1303.0571 [hep-ex]].


\bibitem{Hirose:2016wfn} 
  S.~Hirose {\it et al.} [Belle Collaboration],
  Phys.\ Rev.\ Lett.\  {\bf 118}, no. 21, 211801 (2017)
  [arXiv:1612.00529 [hep-ex]].


\bibitem{Aaij:2015yra} 
  R.~Aaij {\it et al.} [LHCb Collaboration],
  Phys.\ Rev.\ Lett.\  {\bf 115}, no. 11, 111803 (2015)
  Erratum: [Phys.\ Rev.\ Lett.\  {\bf 115}, no. 15, 159901 (2015)]
  [arXiv:1506.08614 [hep-ex]].


\bibitem{Lattice:2015rga} 
  J.~A.~Bailey {\it et al.} [MILC Collaboration],
  Phys.\ Rev.\ D {\bf 92}, no. 3, 034506 (2015)
  [arXiv:1503.07237 [hep-lat]].


\bibitem{Fajfer:2012vx} 
  S.~Fajfer, J.~F.~Kamenik and I.~Nisandzic,
  Phys.\ Rev.\ D {\bf 85}, 094025 (2012)
  [arXiv:1203.2654 [hep-ph]].


\bibitem{Bernlochner:2017jka} 
  F.~U.~Bernlochner, Z.~Ligeti, M.~Papucci and D.~J.~Robinson,
  Phys.\ Rev.\ D {\bf 95}, no. 11, 115008 (2017)
  Erratum: [Phys.\ Rev.\ D {\bf 97}, no. 5, 059902 (2018)]
  [arXiv:1703.05330 [hep-ph]].


\bibitem{Bigi:2017njr} 
  D.~Bigi, P.~Gambino and S.~Schacht,
  Phys.\ Lett.\ B {\bf 769}, 441 (2017)
  [arXiv:1703.06124 [hep-ph]].


\bibitem{Aaij:2017tyk} 
  R.~Aaij {\it et al.} [LHCb Collaboration],
  Phys.\ Rev.\ Lett.\  {\bf 120}, no. 12, 121801 (2018)
  [arXiv:1711.05623 [hep-ex]].


\bibitem{Aaij:2014ora} 
  R.~Aaij {\it et al.} [LHCb Collaboration],
  Phys.\ Rev.\ Lett.\  {\bf 113}, 151601 (2014)
  [arXiv:1406.6482 [hep-ex]].


\bibitem{Aaij:2017vbb} 
  R.~Aaij {\it et al.} [LHCb Collaboration],
  JHEP {\bf 1708}, 055 (2017)
  [arXiv:1705.05802 [hep-ex]].


\bibitem{Hiller:2003js} 
  G.~Hiller and F.~Kruger,
  Phys.\ Rev.\ D {\bf 69}, 074020 (2004)
  [hep-ph/0310219].


\bibitem{Bordone:2016gaq} 
  M.~Bordone, G.~Isidori and A.~Pattori,
  Eur.\ Phys.\ J.\ C {\bf 76}, no. 8, 440 (2016)
  [arXiv:1605.07633 [hep-ph]].


\bibitem{Buttazzo:2017ixm} 
  D.~Buttazzo, A.~Greljo, G.~Isidori and D.~Marzocca,
  JHEP {\bf 1711}, 044 (2017)
  [arXiv:1706.07808 [hep-ph]].


\bibitem{Bhattacharya:2014wla} 
  B.~Bhattacharya, A.~Datta, D.~London and S.~Shivashankara,
  Phys.\ Lett.\ B {\bf 742}, 370 (2015)
  [arXiv:1412.7164 [hep-ph]].


\bibitem{Feruglio:2016gvd} 
  F.~Feruglio, P.~Paradisi and A.~Pattori,
  Phys.\ Rev.\ Lett.\  {\bf 118}, no. 1, 011801 (2017)
  [arXiv:1606.00524 [hep-ph]].

\bibitem{Hiller:2014yaa}
 G.~Hiller and M.~Schmaltz,
  Phys.\ Rev.\ D {\bf 90}, 054014 (2014)
  [arXiv:1408.1627 [hep-ph]].


\bibitem{Assad:2017iib}
  N.~Assad, B.~Fornal and B.~Grinstein,
  Phys.\ Lett.\ B {\bf 777} (2018) 324
  [arXiv:1708.06350 [hep-ph]].
  
\bibitem{Bordone:2018nbg} 
  M.~Bordone, C.~Cornella, J.~Fuentes-Martín and G.~Isidori,
  arXiv:1805.09328 [hep-ph].


\bibitem{Bordone:2017bld} 
  M.~Bordone, C.~Cornella, J.~Fuentes-Martin and G.~Isidori,
  Phys.\ Lett.\ B {\bf 779}, 317 (2018)
  [arXiv:1712.01368 [hep-ph]].


\bibitem{Greljo:2018tuh} 
  A.~Greljo and B.~A.~Stefanek,
  Phys.\ Lett.\ B {\bf 782}, 131 (2018)
  [arXiv:1802.04274 [hep-ph]].


\bibitem{DiLuzio:2017vat} 
  L.~Di Luzio, A.~Greljo and M.~Nardecchia,
  Phys.\ Rev.\ D {\bf 96}, no. 11, 115011 (2017)
  [arXiv:1708.08450 [hep-ph]].


\bibitem{Blanke:2018sro} 
  M.~Blanke and A.~Crivellin,
  arXiv:1801.07256 [hep-ph].


\bibitem{Calibbi:2017qbu} 
  L.~Calibbi, A.~Crivellin and T.~Li,
  arXiv:1709.00692 [hep-ph].


\bibitem{Crivellin:2017zlb} 
  A.~Crivellin, D.~Müller and T.~Ota,
  JHEP {\bf 1709}, 040 (2017)
  [arXiv:1703.09226 [hep-ph]].


\bibitem{Marzocca:2018wcf} 
  D.~Marzocca,
  arXiv:1803.10972 [hep-ph].


\bibitem{Georgi:1974sy} 
  H.~Georgi and S.~L.~Glashow,
  Phys.\ Rev.\ Lett.\  {\bf 32}, 438 (1974).


\bibitem{Hubsch:1984pg} 
  T.~Hubsch and S.~Pallua,
  Phys.\ Lett.\  {\bf 138B}, 279 (1984).


\bibitem{Hubsch:1984qi} 
  T.~Hubsch, S.~Meljanac and S.~Pallua,
  Phys.\ Rev.\ D {\bf 31}, 2958 (1985).


\bibitem{Staub:2013tta} 
  F.~Staub,
  Comput.\ Phys.\ Commun.\  {\bf 185}, 1773 (2014)
  [arXiv:1309.7223 [hep-ph]].

\bibitem{Dorsner:2012nq} 
  I.~Dorsner, S.~Fajfer and N.~Kosnik,
  Phys.\ Rev.\ D {\bf 86}, 015013 (2012)
  [arXiv:1204.0674 [hep-ph]].

\bibitem{Dorsner:2017wwn} 
  I.~Doršner, S.~Fajfer and N.~Košnik,
  Eur.\ Phys.\ J.\ C {\bf 77}, no. 6, 417 (2017)
  [arXiv:1701.08322 [hep-ph]].



\bibitem{Na:2015kha} 
  H.~Na {\it et al.} [HPQCD Collaboration],
  Phys.\ Rev.\ D {\bf 92}, no. 5, 054510 (2015)
  Erratum: [Phys.\ Rev.\ D {\bf 93}, no. 11, 119906 (2016)]
  [arXiv:1505.03925 [hep-lat]].


\bibitem{Lees:2012xj} 
  J.~P.~Lees {\it et al.} [BaBar Collaboration],
  Phys.\ Rev.\ Lett.\  {\bf 109}, 101802 (2012)
  [arXiv:1205.5442 [hep-ex]].


\bibitem{Huschle:2015rga} 
  M.~Huschle {\it et al.} [Belle Collaboration],
  Phys.\ Rev.\ D {\bf 92}, no. 7, 072014 (2015)
  [arXiv:1507.03233 [hep-ex]].


\bibitem{Aaij:2017deq} 
  R.~Aaij {\it et al.} [LHCb Collaboration],
  Phys.\ Rev.\ D {\bf 97}, no. 7, 072013 (2018)
  [arXiv:1711.02505 [hep-ex]].


\bibitem{Amhis:2016xyh} 
  Y.~Amhis {\it et al.} [HFLAV Collaboration],
  Eur.\ Phys.\ J.\ C {\bf 77}, no. 12, 895 (2017)
  [arXiv:1612.07233 [hep-ex]].


\bibitem{Gonzalez-Alonso:2017iyc} 
  M.~González-Alonso, J.~Martin Camalich and K.~Mimouni,
  Phys.\ Lett.\ B {\bf 772}, 777 (2017)
  [arXiv:1706.00410 [hep-ph]].


\bibitem{Dorsner:2017ufx} 
  I.~Doršner, S.~Fajfer, D.~A.~Faroughy and N.~Košnik,
  JHEP {\bf 1710}, 188 (2017)
  [arXiv:1706.07779 [hep-ph]].


\bibitem{Glattauer:2015teq} 
  R.~Glattauer {\it et al.} [Belle Collaboration],
  Phys.\ Rev.\ D {\bf 93}, no. 3, 032006 (2016)
  [arXiv:1510.03657 [hep-ex]].


\bibitem{Abdesselam:2017kjf} 
  A.~Abdesselam {\it et al.} [Belle Collaboration],
  arXiv:1702.01521 [hep-ex].


\bibitem{Becirevic:2016oho} 
  D.~Bečirević, N.~Košnik, O.~Sumensari and R.~Zukanovich Funchal,
  JHEP {\bf 1611}, 035 (2016)
  [arXiv:1608.07583 [hep-ph]].


\bibitem{Cirigliano:2007xi} 
  V.~Cirigliano and I.~Rosell,
  Phys.\ Rev.\ Lett.\  {\bf 99}, 231801 (2007)
  [arXiv:0707.3439 [hep-ph]].


\bibitem{Becirevic:2017jtw} 
  D.~Bečirević and O.~Sumensari,
  JHEP {\bf 1708}, 104 (2017)
  [arXiv:1704.05835 [hep-ph]].


\bibitem{DAmico:2017mtc} 
  G.~D'Amico, M.~Nardecchia, P.~Panci, F.~Sannino, A.~Strumia, R.~Torre and A.~Urbano,
  JHEP {\bf 1709}, 010 (2017)
  [arXiv:1704.05438 [hep-ph]].


\bibitem{Capdevila:2017bsm} 
  B.~Capdevila, A.~Crivellin, S.~Descotes-Genon, J.~Matias and J.~Virto,
  JHEP {\bf 1801}, 093 (2018)
  [arXiv:1704.05340 [hep-ph]].


\bibitem{Altmannshofer:2009ma} 
  W.~Altmannshofer, A.~J.~Buras, D.~M.~Straub and M.~Wick,
  JHEP {\bf 0904}, 022 (2009)
  [arXiv:0902.0160 [hep-ph]].


\bibitem{Grygier:2017tzo} 
  J.~Grygier {\it et al.} [Belle Collaboration],
  Phys.\ Rev.\ D {\bf 96}, no. 9, 091101 (2017)
  Addendum: [Phys.\ Rev.\ D {\bf 97}, no. 9, 099902 (2018)]
  [arXiv:1702.03224 [hep-ex]].


\bibitem{Patrignani:2016xqp} 
  C.~Patrignani {\it et al.} [Particle Data Group],
  Chin.\ Phys.\ C {\bf 40}, no. 10, 100001 (2016).


\bibitem{Beringer:1900zz} 
  J.~Beringer {\it et al.} [Particle Data Group],
  Phys.\ Rev.\ D {\bf 86}, 010001 (2012).


\bibitem{ALEPH:2005ab} 
  S.~Schael {\it et al.} [ALEPH and DELPHI and L3 and OPAL and SLD Collaborations and LEP Electroweak Working Group and SLD Electroweak Group and SLD Heavy Flavour Group],
  Phys.\ Rept.\  {\bf 427}, 257 (2006)
  [hep-ex/0509008].

\bibitem{Sakaki:2014sea}
  Y.~Sakaki, M.~Tanaka, A.~Tayduganov and R.~Watanabe,
  Phys.\ Rev.\ D {\bf 91} (2015) no.11,  114028
  [arXiv:1412.3761 [hep-ph]].
  
\bibitem{Becirevic:2018uab}
  D.~Bečirević, B.~Panes, O.~Sumensari and R.~Zukanovich Funchal,
  JHEP {\bf 1806} (2018) 032
  [arXiv:1803.10112 [hep-ph]].

\bibitem{Lees:2012zz} 
  J.~P.~Lees {\it et al.} [BaBar Collaboration],
  Phys.\ Rev.\ D {\bf 86}, 012004 (2012)
  [arXiv:1204.2852 [hep-ex]].


\bibitem{Becirevic:2016zri} 
  D.~Bečirević, O.~Sumensari and R.~Zukanovich Funchal,
  Eur.\ Phys.\ J.\ C {\bf 76}, no. 3, 134 (2016)
  [arXiv:1602.00881 [hep-ph]].

\bibitem{Glashow:2014iga}
  S.~L.~Glashow, D.~Guadagnoli and K.~Lane,
  Phys.\ Rev.\ Lett.\  {\bf 114} (2015) 091801
  [arXiv:1411.0565 [hep-ph]].

\bibitem{Guadagnoli:2015nra}
  D.~Guadagnoli and K.~Lane,
  Phys.\ Lett.\ B {\bf 751} (2015) 54
  [arXiv:1507.01412 [hep-ph]].
  
\bibitem{Faroughy:2016osc} 
  D.~A.~Faroughy, A.~Greljo and J.~F.~Kamenik,
  Phys.\ Lett.\ B {\bf 764}, 126 (2017)
  [arXiv:1609.07138 [hep-ph]].


\bibitem{Aaboud:2017sjh} 
  M.~Aaboud {\it et al.} [ATLAS Collaboration],
  JHEP {\bf 1801}, 055 (2018)
  [arXiv:1709.07242 [hep-ex]].


\bibitem{Sirunyan:2017yrk} 
  A.~M.~Sirunyan {\it et al.} [CMS Collaboration],
  JHEP {\bf 1707}, 121 (2017)
  [arXiv:1703.03995 [hep-ex]].


\bibitem{CMS:2017kmd} 
  CMS Collaboration [CMS Collaboration],
  CMS-PAS-SUS-16-036.


\bibitem{Sirunyan:2018nkj} 
  A.~M.~Sirunyan {\it et al.} [CMS Collaboration],
  arXiv:1803.02864 [hep-ex].
\end{thebibliography}
\end{document}